\documentclass[1p]{elsarticle}
\usepackage{aas_macros}
\usepackage{url}
\newcommand{\sminy}{\fontsize{9}{11}\selectfont}
\usepackage{array}
\usepackage{amsmath}
\usepackage{cite}
\begin{document}

\begin{frontmatter}
\title{Efficient Parallelization for AMR MHD Multiphysics Calculations; Implementation in AstroBEAR}

\author[1]{Jonathan J. Carroll-Nellenback}
\ead{johannjc@pas.rochester.edu}
\author[1]{Brandon Shroyer}
\author[1]{Adam Frank}
\author[2]{Chen Ding}
\address[1]{Department of Physics and Astronomy, University of Rochester, Rochester, NY 14620}
\address[2]{Department of Computer Science, University of Rochester, Rochester, NY 14620}

\begin{abstract}
Current Adaptive Mesh Refinement (AMR) simulations require algorithms that are highly parallelized and manage memory efficiently.  As compute engines grow larger, AMR simulations will require algorithms that achieve new levels of efficient parallelization and memory management.  We have attempted to employ new techniques to achieve both of these goals.  Patch or grid based AMR often employs ghost cells to decouple the hyperbolic advances of each grid on a given refinement level.  This decoupling allows each grid to be advanced independently.  In AstroBEAR we utilize this independence by threading the grid advances on each level with preference going to the finer level grids.  This allows for global load balancing instead of level by level load balancing and allows for greater parallelization across both physical space and AMR level.  Threading of level advances can also improve performance by interleaving communication with computation, especially in deep simulations with many levels of refinement.  While we see improvements of up to $30 \%$ on deep simulations run on a few cores, the speedup is typically more modest ($5-20 \%$) for larger scale simulations.  To improve memory management we have employed a distributed tree algorithm that requires processors to only store and communicate local sections of the AMR tree structure with neighboring processors.  Using this distributed approach we are able to get reasonable scaling efficiency ($ > 80 \%$) out to 12288 cores and up to 8 levels of AMR - independent of the use of threading.
%We have also implemented a sweep method that pipe-lines the computations required for updating the fluid variables using unsplit algorithms.  This can dramatically reduce the memory overhead required for intermediate variables.
\end{abstract}

%\keywords{protostellar outflows, turbulence, star formation rate}

\end{frontmatter}

\section{Introduction}

The development of AMR \citep{Berger1984484, Berger198964, Balsara2001} was meant to provide high resolution simulations for much lower computational cost than fixed grid methods would allow.  The use of highly parallel systems and the algorithms that go with them were also meant to allow higher resolution simulations to be run faster (relative to wall clock time).  The parallelization of AMR algorithms, which should combine the cost/time savings of both methods is not straight forward however and there have been many different approaches \citep{Paramesh2000, Nirvana2008, Enzo2004, Chombo, FTT} .  While parallelization of a uniform mesh demands little communication between processors, AMR methods can demand considerable communication to maintain data consistency across the unstructured mesh as well as shuffling new grids from one processor to another to balance workload.  

In this paper we report the development and implementation of new algorithms for the efficient parallelization of AMR designed to scale to very large simulations. The new algorithms are part of the AstroBEAR package for simulation of astrophysical fluid multi-physics problems \citep{Cunningham2009ApJS}.  The new algorithmic structure described in this paper constitutes the development of version 2.0 of the AstroBEAR code. 

AMR methods come in many varieties.  Meshes can either be unstructured or semi-structured.  Semi-structured methods can be further divided into those which allow grids to be of arbitrary size (patch based) and those which require grids to be of a fixed size (block based or cell-based if the block size is 1).  With block (or cell) based AMR, the additional constraints imposed on the structure of the mesh allow for a simpler type of connectivity within a tree.  For example in block based AMR, any given block will have exactly 8 children or none (if it is a leaf) and will have at most 6 face sharing neighbors.  With patch based AMR, there is no limit to the number of children or neighbors.  In addition, the operation of regridding in block based AMR is much simpler.  As the grid changes a given block will either persist if the physical region continues to require refinement or be destroyed. In patch based AMR, a given region may subsequently be better covered by patches of a different shape requiring transfer of data between physically overlapping previous patches and new patches.  This adds an additional dimension to the tree structure and increases the complexity of maintaining a distributed tree.

For both block (or cell) and patch based AMR, the actual grid data (fluid variables etc.) are always distributed across the various processors.  Usually some overlap in grid data (guard/ghost cells) is desired to allow for frequent access to neighboring values without the need for additional communication.  But the metadata that describes the shape and distribution of the grid data is usually stored on every processor.  For 100's or 1000's of cores, this global tree typically requires less memory than that required for the local grid data and it allows for easy access to data from any part of the domain.  For instance after regridding, the entire tree can be updated and stored locally and then finding new neighbors for local cells/blocks/patches can be done without the need for any further communication.  For patch-based AMR it also allows for 'better' load balancing as each processor can determine from the entire tree which section of data it should be responsible for and can use various knap-sack type algorithms to optimize the degree of interprocessor communication.

Cell based AMR engines typically use an Octree data structure to handle the tree metadata and implementations have been developed that support a memory-distributed tree \citep{DistributedOctTree, Octor}, or various ways of compressing the global tree \citep{CompressedOctTrees, FTT} that rely on the simple structure of the tree.  For patch based AMR, algorithms for implementing a distributed tree have not yet been published though implemented in the package SAMRAI \citep{Chombo} as well as AstroBEAR.  In addition the Chombo library \citep{Chombo} recently has developed a method of compressing the metadata to avoid having to distribute the tree.

Here we document the distributed tree algorithm used in AstroBEAR 2.0 in which no processor has access to the entire tree but rather each processor is only aware of the AMR structure it needs to manage in order to carry out its computations and perform the necessary communications.  While currently, this additional memory is small compared to the resources typically available to a CPU, future clusters will likely have much less memory per processor similar to what is already seen in GPU's.  Additionally each processor only sends and receives the portions of the tree necessary to carry out its communication as opposed to a pruning approach in which every processor receives every new patch, but only keeps those necessary to maintain the local tree.

AstroBEAR 2.0 also uses extended ghost cells to decouple advances on various levels of refinement.  As we show below this allows for each level's advance to be computed independently on separate threads.  Such inter-level threading allows for total load balancing across all refinement levels instead of balancing each level independently.   Independent load balancing becomes especially important for deep simulations (simulations with low filling fractions but many levels of AMR) as opposed to shallow simulations (high filling fractions and only a few levels of AMR).  Processors with coarse grids can advance their grids simultaneously while processors with finer grids advance theirs.  Without such a capability, each level would need to have enough cells to be able to be distributed across all of the processors.  Variations in the filling fractions from level to level can make the number of cells on each level very different.  If there are enough cells on the level with the fewest to be adequately distributed, there will likely be far too many cells on the highest level to allow the computation to be completed in a reasonable wall clock time.  This often restricts the number of levels of AMR that can be practically used.  With inter-level threading this restriction is lifted.  Inter-level threading also allows processors to remain busy while waiting for messages from other processors.

In what follows we provide descriptions of the new code and its structure as well as providing tests which demonstrate its effective scaling.  In section~\ref{amr_alg} we review patch based AMR.  In section~\ref{distributedtree} we describe in detailt the distributed tree algorithm for patch-based AMR, in section~\ref{threading} we will discuss the inter-level threading of the advance, in section~\ref{loadbalancing} we will discuss the load balancing algorithm, and in section~\ref{results} we will present our scaling results and we will conclude in section~\ref{conclusion}.

% and in section~\ref{sweep} we will discuss the pipe-lining of the unsplit integration schemes.  We will also briefly discuss attempts to further reduce locally redundant advance computations in section~\ref{supergridding} 

\section{AMR Algorithm}\label{amr_alg}

\begin{figure}
 \caption{Example AMR mesh showing nested and adjacent grids as well as corresponding AMR tree showing parent-child and neighbor relationships.}
 \centering
  \includegraphics[width=1.0\textwidth]{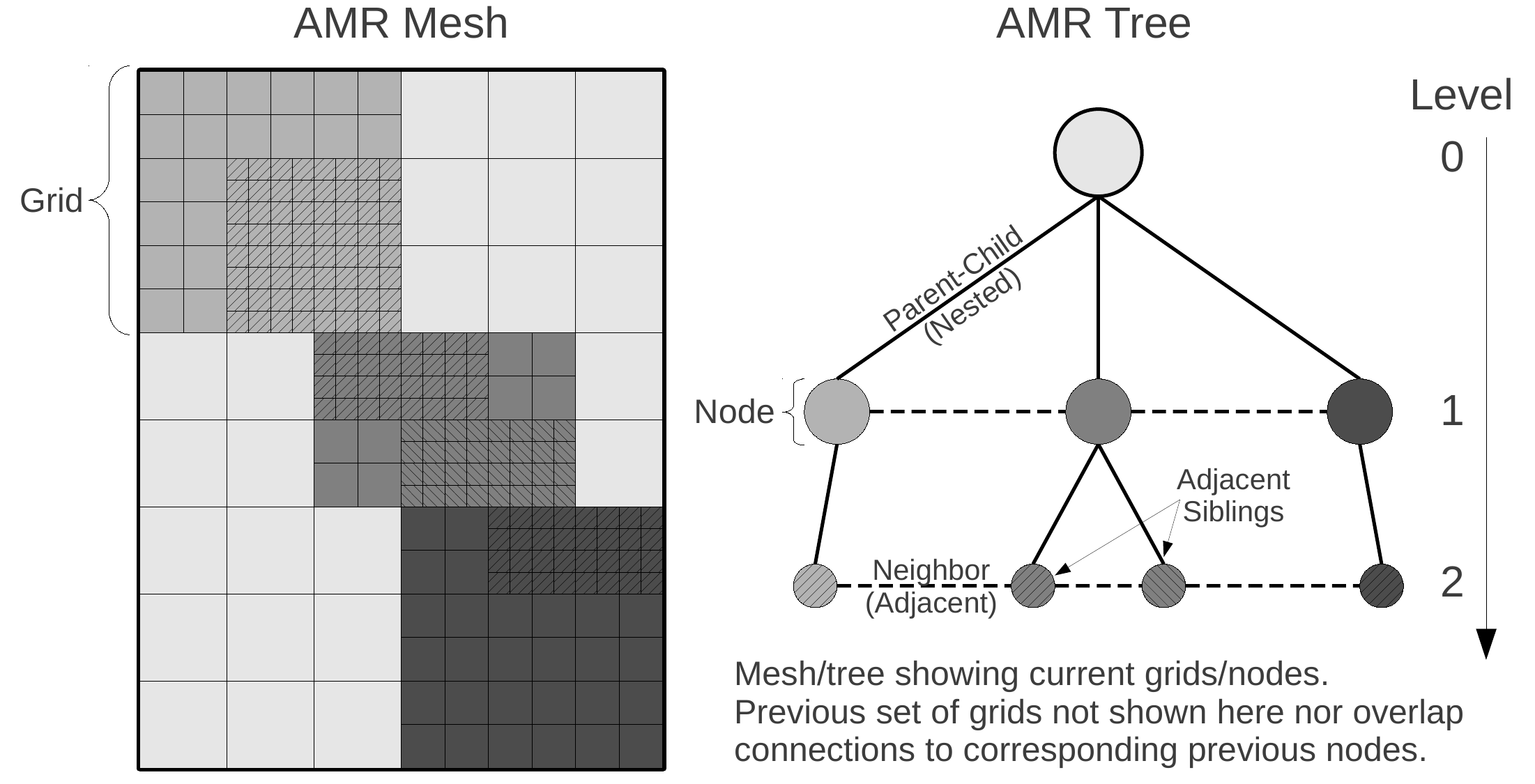} 
\label{meshtree}
\end{figure}

  Here we give a brief overview of patch based AMR introducing our terminology along the way.  The fundamental unit of the AMR algorithm is a patch or grid.  Each grid contains a regular array of cells in which the fluid variables are stored.  Grids with a common resolution or cell width $\Delta x_l$ belong to the same level $l$ and on all but the coarsest level are always nested within a coarser ``parent" grid of level $l-1$ and resolution $\Delta x_{l-1} = R \times \Delta x_l $ where $R$ is the refinement ratio.  The collection of grids comprises the AMR mesh, an example of which is shown in figure~\ref{meshtree}.  In addition to the computations required to advance the fluid variables, each grid needs to exchange data with its parent grid (on level $l-1$) as well as any child grids (on level $l+1$).  Grids also need to exchange data with physically adjacent neighboring grids (on level $l$).  In order to exchange data, the physical connections between parents, children, and neighboring grids are stored in the AMR tree as connections between nodes.  Each grid has a corresponding node in the AMR tree.  Thus there is a one to one correspondence between nodes and grids.  The grids hold the actual fluid dynamical data while the nodes hold the information about each grid's position and its connections to parents, children and neighbors.  Figure~\ref{meshtree} shows one example of an AMR mesh made of grids and the corresponding AMR tree made of nodes.  Note that what matters in terms of connections between nodes is the physical proximity of their respective grids.  While siblings share a common parent, they will not necessarily be neighbors, and neighbors are not always siblings but may be 1st cousins, 2nd cousins, etc...  

\begin{figure}
 \caption{Example AMR mesh showing two iterations of level 1 grids and the overlapping regions.  Nodes associated with previous/successive iterations of overlapping grids will have preceding/succeeding overlap connections in the AMR tree.}
 \centering
  \includegraphics[width=1.0\textwidth]{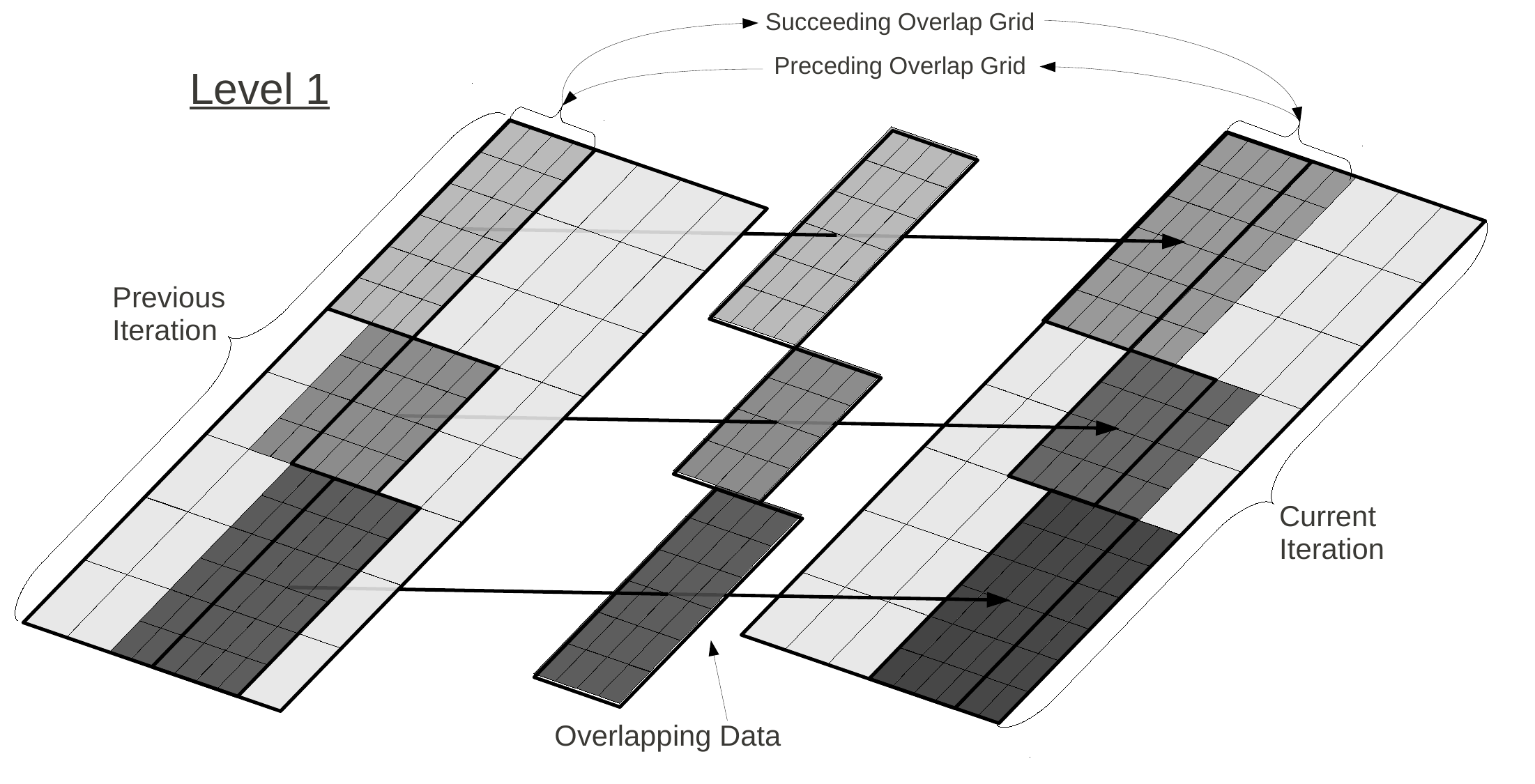} 
\label{overlaps}
\end{figure}

  Additionally since the mesh is adaptive there will be successive iterations of grids on each level as the simulation progresses.  Thus the fluid variables need to be transferred from the previous iteration of grids to the current iteration  as shown in figure~\ref{overlaps}.  Thus nodes can have ``neighbors'' within a 4 dimensional space-time.  Nodes that are temporally adjacent (belonging to either the previous or next iteration) and spatially coincident are classified as preceding or succeeding overlaps respectively instead of temporal neighbors, reserving the term neighbor to refer to nodes of the same iteration that are spatially adjacent and temporally coincident.  Nodes on level $l$ therefore have a parent connection to a node on level $l-1$, child connections to nodes on level $l+1$, neighbor connections to nodes on level $l$ of the same iteration, and overlap connections to nodes on level $l$ of the previous or successive iteration in time.

\subsection{Regridding}
We use the term ``iteration" to refer to successive generations of the grid distribution for each level.  Thus at some point in the simulation a distribution of level $l$ grids will cover some fraction of the computational space.  When, as the simulation proceeds, the grid generation subroutine is called again the next iteration of level $l$ grids will be laid down over the computational domain.  

While in principle level $l$ could be regridded each level $l$ time step, it is convenient for restricting data to parent grids, to wait until after $R$ level $l$ steps, or equivalently after $1$ parent level $(l-1)$ step so that child and parent grids have advanced to the same time.  Therefore for each level $0$ time step, there is $1$ iteration of level $1$ grids, $R$ iterations of level $2$ grids, $R^2$ iterations of level $3$ grids, and so on.  Since new child grids are created after each parent step, each parent will have multiple iterations of children.  This additional branching of the AMR tree in time allows for temporally adjacent (overlap) grids to be classified as temporal siblings or 1st, 2nd, 3rd temporal cousins etc...  Additionally parents will have connections to multiple iterations of children (although only the two most current need to be kept).

Note that there are different ways of indexing the iteration of grids on a given level. One can chose a global indexing that begins with the initiation of the simulation, i.e the iterations can be indexed by counting the number of successive iterations of grids on that level from the beginning of the simulation.  Another method is to index iterations relative only to the level above (IE which iteration of children from a given parent does a grid correspond to.)  We will use the latter indexing scheme, since it is relevant to the way overlap connections between grids of different iterations are formed.  

\begin{figure}
 \caption{Time dependence of AMR tree showing overlaps between old grids and new grids.  Each node is labeled by the iteration of its parent's children.}
 \centering
  \includegraphics[width=1.0\textwidth]{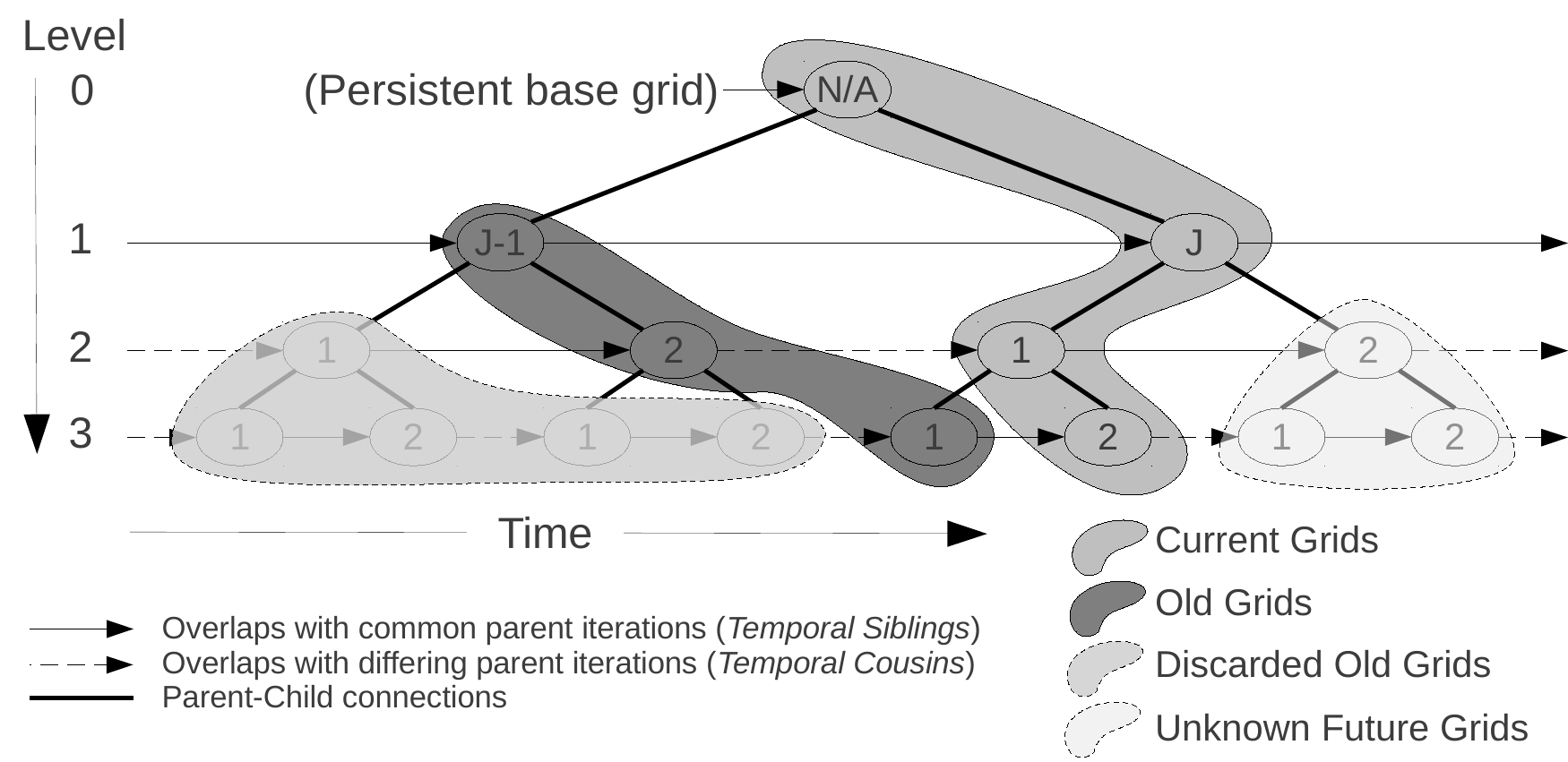} 
\label{overlapfigure}
\end{figure}

  Figure~\ref{overlapfigure} shows this indexing system (where we have assumed $R=2$ for simplicity).  Here the circles represent iterations of the entire level (i.e. the entire collection of nodes for a given level at the specified iteration).    Since figure~\ref{overlapfigure} shows the evolution of the AMR tree {\it in time}, it's useful to make a connection with the the way the tree appears at any given moment {\it in space} in terms of spatial connections between nodes (figure~\ref{meshtree}).  To understand the relation between figure~\ref{meshtree} and figure~\ref{overlapfigure} imagine taking the tree in figure~\ref{meshtree} and rotating it into the plane.  Neighbor relationships would now disappear and all of the nodes of a given level would visually merge.  Each visible circle in figure~\ref{overlapfigure} represents a collection of nodes of the same level and iteration as shown in figure~\ref{meshtree}.  Each level $(l>0)$ has multiple iterations that stretch both forward and backward in time.  Note that the level 0 grids are static and there are no successive iterations.  Level 1 iterations go from $1$ to $\infty$ since the level 0 grids continue to create successive iterations of children.  Level 2 iterations and higher are indexed from $1$ to $R=2$.  
  
 It is important to stress here, that each level's iteration in figure~\ref{overlapfigure} represents a collection of nodes on that level.  Additionally each level's iteration represents multiple time steps on that level.  The nodes of the level 0 iteration actually take many time steps, while all other level iterations take $R=2$ time steps.  Preceding overlap connections (those going backward in time) are only needed before the 1st of $R$ time steps, and the succeeding overlap connections (those going forward in time) are only needed after the last of $R$ time steps.  After the 1st time step and before the last, ghost overlap data is shared between neighboring grids of the same iteration instead of preceding/succeeding overlaps.  

While there will be many iterations of grids on each level, only the two most current iterations are stored in the tree.  After a grid finishes its advance it becomes ``old".  When this occurs the previous generation of old grid iterations are discarded.  The current and old grids are shown schematically in figure~\ref{overlapfigure}.  The current level $l$ iteration will always contain children of the current level $l-1$ iteration. For the old level $l$ iteration there are two cases.  First the old level $l$ iteration may be the old iteration of children of the current $l-1$ iteration as is the case for the old level 3 grids in figure~\ref{overlapfigure}.  Secondly, the old level $l$ iteration may be the last iteration of children of the old level $l-1$ iteration as is the case for the old level 2 grids in figure~\ref{overlapfigure}.  That is, they are either old children of the current parent level grids, or last children of the old parent level grids.

\subsection{Grid Life Cycle}

The execution of the various stages in the AMR algorithm is shown in the upper panel of Figure~\ref{threadingfig} (for $R=2$). There are 5 steps in the algorithm: overlaps (O); Prolongation (P); Advances (A); Synchronization (S) and Restriction (R).  A pseudo-code description of their order of operations with $L$ being the highest level of refinement is written below.  Note each root step is initialized by calling AMR($0$).

\indent RECURSIVE SUBROUTINE AMR($l$)

\indent \indent \textbf{O}verlap data from preceding iterations of grids on level $l$

\indent \indent DO $i$=1, $R$

\indent \indent \indent \textbf{A}dvance grids on level $l$

\indent \indent \indent IF ($l < L$) THEN

\indent \indent \indent \indent \textbf{P}rolongate initial (pre-advanced) data from level $l$ to level $l+1$

\indent \indent \indent \indent CALL AMR($l+1$)

\indent \indent \indent \indent \textbf{R}estrict data from level $l+1$

\indent \indent \indent END IF

\indent \indent \indent \textbf{S}ynchronize fluxes between neighboring level $l$ grids and update ghost cells

\indent \indent END DO

\indent END RECURSIVE SUBROUTINE

The subroutine above describes a process in which grids are created, populated with data, advanced and brought in line with higher and lower refinement representations of the data.  To be more specific, after each iteration of level $l$ grids are created their cells (including ghost regions) are initialized  with a combination of prolongated data from the parent grid as well as data from the preceding set of level $l$ grids that physically overlap with the computational space that the new grids now describe.  Ghost regions are needed to update the fluid variables within the grid.  The new level $l$ grids then determine which cells to refine and lay down a first set of level $l+1$ child grids to cover those cells.  The parent level $l$ grids take one step of $\Delta t_l$.  Meanwhile their child grids advance $R$ steps of $\Delta t_{l+1} = \Delta t_{l}/R$.  Next the level $l$ grids merge restricted data from their children with their own updated data since both levels have now reached the same time $t$.  These level $l$ grids then synchronize fluxes with any adjacent neighboring grids (also of level $l$) and update any ghost cells.  The level $l$ grids are now ready to repeat the process (creating children; advance a time step; restricted child data; synchronize fluxes with neighbors).  When the level $l$ grids have completed $R$ steps relative to the level above ($l-1$), their own data and fluxes are restricted and applied to their parent ($l-1$) grids.  

After completing its advances, the data from each level $l$ grid is stored until it can be copied onto the next iteration of level $l$ grids.  After the data is copied onto {\it succeeding} overlaps, the old level $l$ grids are destroyed.  Throughout a level $l$ grid's lifetime it must therefore share data with its parent grid $(l-1)$, its preceding overlaps ($l$), multiple iterations of it own child grids ($l+1)$, its neighbor grids  $(l)$ , and succeeding overlaps  $(l)$.  These connections for a given grid are held by its node and the web of connections between grids/nodes form the AMR tree.  While a grid/node may have many successive iterations of children, only connections that belong to the two most current iterations of the child level are kept.

\subsection{Isolated Grids}
\begin{figure}
 \caption{Two solutions for isolated grids.  The one on the right requires additional calculations to update the ghost cells once, but allows for each level of the AMR hierarchy to advance independently.}
 \centering
  \includegraphics[width=1.0\textwidth]{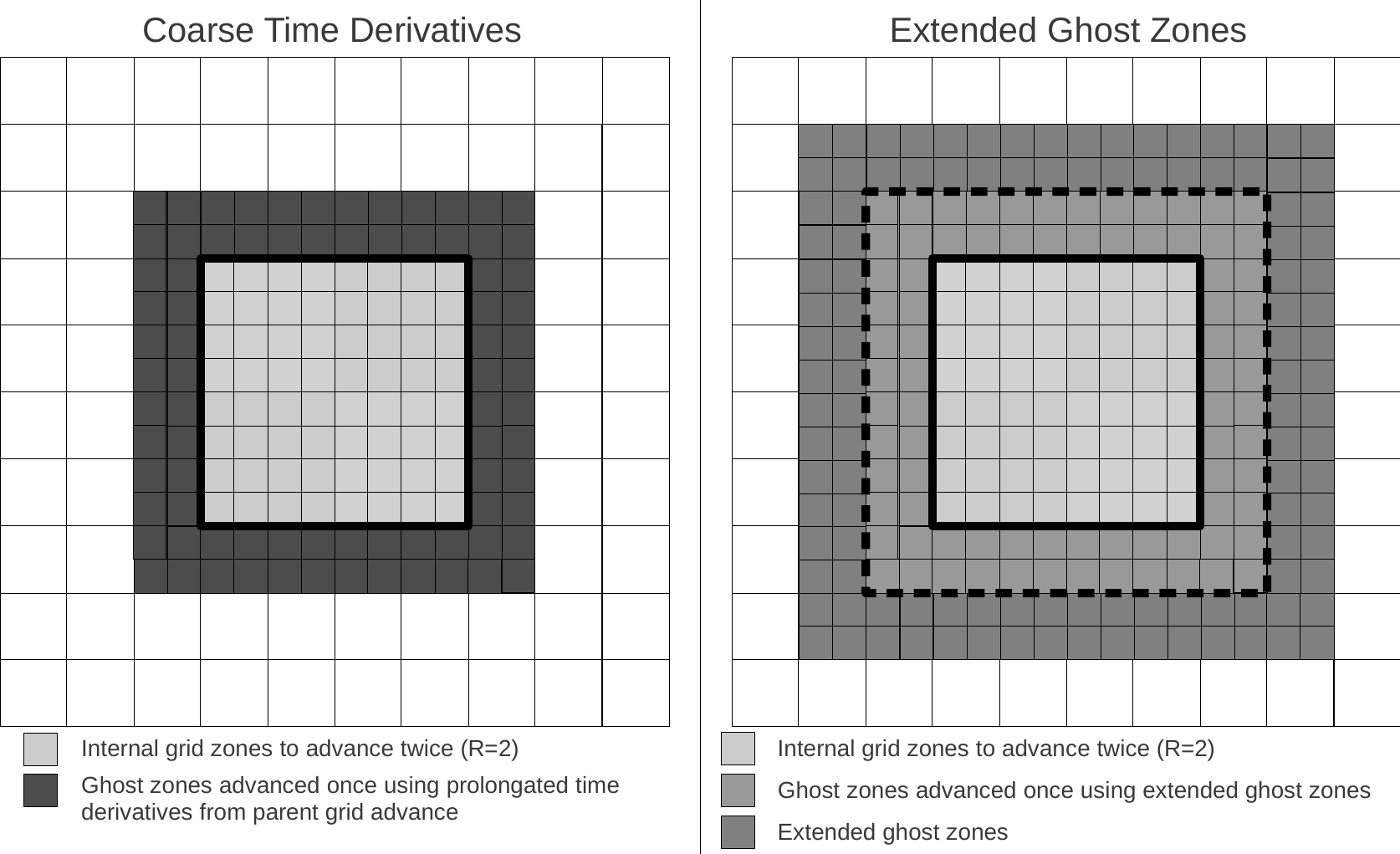} 
\label{extendedghostcells}
\end{figure}

  Refined grids need to take multiple ($R$) time steps.  For the first step, prolongated ghost cells from the coarser grid can be used to advance the internal cells.  However further time steps cannot be taken without updating the ghost cells in time as well.  If a grid is surrounded by neighboring grids, it can update its ghost cells in time by copying the advanced data from its neighbors' internal cells.  However, if a grid is isolated (ie. has no neighbors) or even if it is partially isolated, it must advance those ghost cells in another way.  There are two solutions to this problem shown in figure~\ref{extendedghostcells}.  One is to use the time derivative for the fluid variables calculated on the coarse grid to update the ghost cells where needed.  This solution requires parent grids to advance before their children and can lead to large errors if shocks are not completely resolved since spatial discontinuities lead to delta functions for time derivatives which, in turn, produce large errors upon discretization.  

The alternative method is to use extended ghost cells that allow grids to successively update smaller and smaller regions so that there is always enough ghost cells to complete the final step.  In general if $n_{\mbox{\small{ghost}}}$ is the number of ghost cells needed to take a time step and $R$ is the refinement ratio, then the extended ghost region needs to initially be $R\times n_{\mbox{\small{ghost}}}$ cells wide.  While only isolated grids need to update extended ghost cells, it can be difficult to implement efficient advance schemes for partially isolated grids, and each grid is often assumed to be isolated.  While this can result in a large overhead for small grids, (especially when the refinement ratio $R$, or $n_{\mbox{\small{ghost}}}$ is large) the ability to allow coarse grids to be advanced independently of fine grids can be exploited to increase the speed of the code. We have taken this approach in AstroBEAR to allow for more efficient distribution of grids as well as the creation of multiple advance threads described in section~\ref{threading}.

\section{Distributed Tree Algorithm}\label{distributedtree}

Many current AMR codes store the entire AMR tree on each processor.  This, however, can become a problem for simulations run on many processors.  To demonstrate this let us first assume that each AMR grid requires $m$ bytes per node to store its meta data (ie 6 bytes to store its physical bounds for a 3D simulation and 1 byte to store the processor containing the grid).  We also assume that each grid requires on average $d$ bytes for the actual data.  If there are, on average, $n$ grids on each of $p$ processors, then the memory per processor would be $nd + nmp$.  The second term $nmp$ represents the meta-data for the {\it entire} AMR tree.  

The memory requirement for just the nodes in the AMR tree without storing any connections becomes comparable to the local actual data when $p = d/m$.  If we assume a 3D isothermal hydro run where each cell contains $\rho, p_x, p_y, \& p_z$ with a typical average grid size of 8x8x8 then $p=\frac{8\times8\times8\times4}{(6+1)} \approx 293$.  While this additional memory requirement is negligible for problems run using typical CPU's on 100's of processors, it can become considerable when the number of processors $n > 10^3$.  Since it is expected that both efficient memory use and management will be required for ever larger HPC (high performance computing) clusters down the road, AstroBEAR 2.0 is designed to use a {\it distributed} tree algorithm. In this scheme each processor is only aware of the section of the tree containing nodes that connect to its own grids' nodes.  Additionally, new nodes are communicated to other processors on a proscriptive ``might need to know basis''.  Since maintaining these local trees as the mesh adapts is not trivial, we describe the process below.

\subsection{Maintaining the AMR Tree}

 Because of the nested nature of grids, neighbor and overlap relationships between nodes can always be inherited from parent relationships.  First consider the neighbors of the $n^{th}$ iteration of a node's children.  The nested nature of the grids restricts each of the child's neighbors to either be a sibling of that child (having the same parent node and be of the same iteration), or to be a member of a neighbor's $n^{th}$ iteration of children.  Thus the neighbors of a level $l$ node's children (on level $l+1$) will always be a child of that level $l$ node's neighbors.  

  For overlaps (temporal neighbors) the situation is a bit more complicated.  If we consider figure~\ref{overlapfigure} we can identify two different types of overlap connections.  First there are overlap connections between iterations with a common parent which we will refer to as 'temporal siblings'.  For example the current and old iterations shown on level 3 are different iterations of children of the same level 2 iteration.  There are also overlap connections between children that come from different iterations of the parent grids.  These can be thought of as 'temporal cousins'.  For example the current and old child iterations shown on level 2 come from different iterations of the parent nodes one level above them.  In principle  temporal cousins can be 1st cousins, 2nd cousins, and so on, but the distinction does not matter for the way connectivity is treated in the algorithm.  

  Overlaps between temporal siblings on level $l$ will be between successive child iterations of the current level $l-1$ iteration.  Note that grids in the current $l-1$ iteration do not physically overlap.   Since grids are always nested, the overlaps between the grid's current ($n^{th}$) iteration of children must belong to the grid's old ($n^{th}-1$) iteration of children (for completeness these are what are referred to as preceding overlaps).  However, while grids do not physically overlap, their ghost cells may. Thus overlapping data for a level $l$ grid's children may also come from a neighbor level $l$ grid's old children.  Thus the (preceding) overlaps for a node's children may be an old child of the node's neighbors.  Likewise, every node's old child's succeeding overlap may be a node's neighbor's current child.

  For temporal cousins, the current iteration of level $l-1$ grids/nodes will be the first iteration of children of the current level $l$ grids/nodes.  And the old iteration of level $l-1$ grids will be the last iteration of children of the old level $l$ grids.  Because of the nested nature of the grids, we can say that for the first iteration of children, every node's child's preceding overlap must be a node's preceding overlap's child.  The same follows for the succeeding overlaps of a node's last iteration of children. That is, every node's child's succeeding overlap must be a node's succeeding overlap's child.  These relationships are summarized for $R=2$ in table \ref{simpleinheritance} and a generalized table is given in table \ref{geninheritance}.

\begin{table}\footnotesize
\centering
\begin{tabular}{|>{\centering}p{.82in}|p{.9in}|p{1.15in}|p{1.1in}|}
\hline
 Child iteration  & Node's child's neighbors & Node's child's preceding overlaps & Node's child's succeeding overlaps\\
\hline
 1$^{st}$ & Node's neighbors' 1$^{st}$ children & Node's preceding overlaps' 2nd children & Node's neighbor's 2$^{nd}$ children  \\
\hline
 2$^{nd}$ & Node's neighbors' 2$^{nd}$ children & Node's neighbors' 1$^{st}$ children & Node's succeeding overlaps' 1$^{st}$ children \\
\hline
\end{tabular}
\caption{Simplified inheritance pattern for a refinement ratio of 2}
\label{simpleinheritance}
\end{table}

\subsection{Maintaining Distributed Sub-Trees}
  For parallel applications, the grids are distributed across the processors.  In addition to data for the local grids, each processor needs to know where to send and receive data for the parents, neighbors, overlaps, and children of those local grids.  This is information contained within the nodes.  In order for a processor to know where to send data, each one must maintain a local sub-tree containing its own ``local" nodes ( corresponding to local grids) as well as all remote nodes (living on other processors) directly connected to the local nodes.  It is also possible, though not desirable, that an individual processor have data from disjoint regions of the simulation.  In that case each processor would have multiple disjoint sections of the AMR tree, but these disjoint sections would collectively be considered the processor's sub-tree.

Each time new grids on level $l+1$ are created (by local parent grids on level $l$), each processor determines how the new child grids should be distributed (i.e. which processor should get the new grids).  This distribution is carried out in the manner described in section~\ref{loadbalancing} below.  Connections between the new level $l+1$ nodes and the rest of the tree must then be formed.  Because of the inheritability of the neighbor/overlap connections, even if a child grid is distributed to another processor, the connections between that child node and its neighbors/overlaps/parent are first made on the processor that created the grid (ie the processor containing the grid's parent).  If a processor's local grid has a remote parent, then that processor will always receive information about that local grid's neighbors/overlaps from the processor containing the remote parent.  This is true of neighbor and preceding overlap connections of new grids as well as succeeding overlap connections of old grids.

Before processors containing parents of level $l+1$ grids (both new and old) can send connection information to remote children (if they exist), these processors must first share information about the creation of children with each other.  Neighbor connections between new nodes on level $l+1$ require each processor to cycle through its local level $l$ nodes and identify those with remote neighbors living on other processors.  Once remote neighbors have been identified the information about new children from the local nodes is sent to the processor(s) containing the remote neighbors.  Not all children need to be sent to every remote neighbor.  Only those that are close enough to potentially be adjacent to the remote neighbor's children are necessary.  The information must flow in both directions meaning a individual processor also needs to {\it receive} information about potential new children from all other processors containing remote neighbors.

Overlaps are again a bit more complicated because of the different situations for temporal siblings and temporal cousins as described above.  For temporal cousins (ie if the new level $l+1$ child grids are the 1st iteration from level $l$ parents), each processor must cycle through local current level $l$ nodes with preceding overlaps that live on remote processors.  The processor must then send new children of local nodes to the processor(s) containing the remote preceding overlaps.   Similarly, each processor must cycle through the local \textbf{old} level $l$ nodes with remote succeeding overlaps sending children of those old local nodes to the processor(s) containing remote succeeding overlaps.  Again, not all children need to be sent to the processors containing remote overlaps. Only information about those nodes close enough to potentially overlap the remote node's children must be sent.  The information must flow in both directions and processors also need to receive information about potential children from those same processors with remote overlaps.

For overlaps between temporal siblings, no communication is required.  This is because previous level $l+1$ nodes that could overlap a new level $l+1$ child node would either be old children of the same parent or be old children of the parent's neighbors.  The parent would already be aware of its old children, and the old children of the parent's neighbors would have been previously communicated when establishing the neighbor relationships of the parent's old children.  This process is summarized in table \ref{DistTreeAlg}.

\section{Threaded Multilevel Advance} \label{threading}

\begin{figure}
 \caption{Plot showing threads of AMR algorithm}
 \centering
  \includegraphics[angle=90]{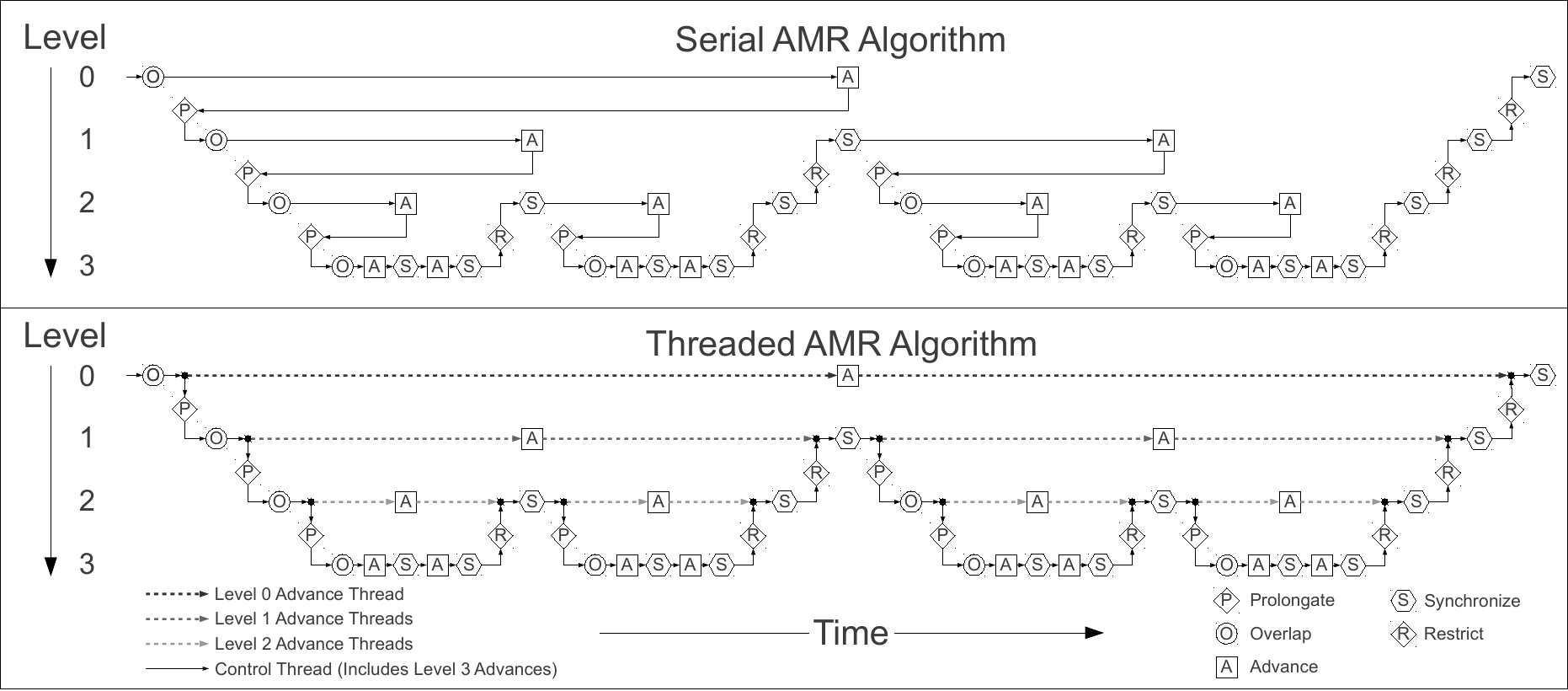} 
\label{threadingfig}
\end{figure}

 Many if not all current AMR codes tend to perform grid updates across all levels in a prescribed order that traverses the levels of the AMR hierarchy in a sequential manner.  Thus the code begins at the base grid (level 0), moves down to the highest refinement level and then cycles up and down across levels based on time step and synchronization requirements(for a simulation with 3 levels the sequence would be: 0, 1, 2, 2, 1, 2, 2, 0...) In the top panel of figure~\ref{threadingfig} the basic operations of (P)rolongating, (O)verlapping, (A)dvancing, (S)ynchronizing, and (R)estricting are shown for each level along with the single (serial) control thread.  
 
 Good parallel performance requires each level update to be independently balanced across all processors (or at least levels with a significant fraction of the workload).  Load balancing each level, however, requires the levels to contain enough grids to be effectively distributed among the processors.  Such a requirement demands each level to be fairly ``large" in the sense of having many grids or allowing each level's spatial coverage be artificially fragmented into small pieces.  The former situation leads to broad simulations (large base grid leaving resources for only a few levels of AMR), while the later situation leads to inefficient simulations due to the fair amount of overhead required for ghost cell calculations.

The problem becomes worse when $n_{ghost}$ is large or when using extended ghost cells with refinement ratios $ R > 2$.  Consider an isolated level $l$ grid of size $4 \times 4$ inside of a parent grid on level $l$.  Let's assume that the coarsening ratio $R=2$ and that the grid must therefore take two steps each of which requires $n_{ghost}=4$ ghost cells.  On the first step it must update a region that is $12 \times 12$ and then on the second step a region that is $4 \times 4$ for a total of $12 \times 12 + 4 \times 4 = 160$ cell updates.  If that $4 \times 4$ grid were split into 4 $2 \times 2$ pieces to be distributed on 4 processors, then instead of 160 cell updates there would now be $10 \times 10 + 2 \times 2 = 104$ per processor or 416 total cell updates and the parallelization efficiency for that grid would be $160/416=38\%$ at best.  Even if one uses prolongated time derivatives instead of extended ghost cells, there is still a fair amount of overhead for ghost cell computations for small grids and the efficiencies would still be $\approx 60-70\%$ depending on the particular method used.

In the bottom panel of figure~\ref{threadingfig} we show a schematic of the AstroBEAR 2.0 AMR algorithm.  In this figure~basic operations of (P)rolongating, (O)verlapping, (A)dvancing, (S)ynchronizing, and (R)estricting  are shown again however this time the level advances are independent and exist on separate threads of computation.  There is an overarching control thread which handles all of the communications and computations required for prolongation, overlapping, synchronizing, and restricting as well as the finest level advances.  Each coarser level advance has its own thread and can be carried forward independently with preference being given to the threads that must finish first (which is always the finer level threads).  In addition to relaxing the requirement of balancing every level, the existence of multiple threads allows processors to remain busy when the control thread becomes held up because it needs information from another processor.  For example, while waiting for ghost cell data for level 3 it can work on advancing levels 2, 1, or 0.

The simplest implementation of threading would allow for the operating system kernel to manage the various threads with priority given to the control thread followed by the highest level advance and so on all on the same processor.  This would be considered preemptive prioritized threading within a process.  Unfortunately the Pthreads library implementation under Linux does not allow for non-privileged users to give priorities to threads nor to limit a set of threads to use only a single processor.  GNU portable threads do require the threads to remain on a single processor, but do not allow for preemptive thread scheduling.  Threads must schedule themselves cooperatively by yielding control to the scheduler or to each other. While GNU portable threads do allow for priorities to be assigned to threads, it is not terribly useful as threads can cooperatively yield control to the appropriate thread without the need for a scheduler.  A more complicated implementation, but one that does not require external libraries, is to manually switch (at the application level) between grid advances and the control thread.  This is somewhat difficult to implement as the grid advances must be interruptable and any intermediate variables within the advance stack have to be manually cached by the application.  This caching effectively limits the frequency with which the advance thread can be interrupted.  In AstroBEAR 2.0 we have implemented GNU portable threads, manual thread switching (scheduling), as well as the serial version of the AMR algorithm. Performance results are given in section~\ref{results}.

%\begin{figure}
% \caption{Sample Distributions for level by level load balancing vs global load balancing on 4 processors.}
% \centering
 % \includegraphics{IdealDistribution.pdf} 
%\label{idealizedcase}
%\end{figure}

%\includegraphics[width=4in]{figure1.png}
%\includegraphics[width=5in]{figure1.eps} \label{threadingfig}

\section{Load Balancing} \label{loadbalancing}
As was discussed above, threading level advances removes the need for balancing each level independently and instead allows for global load balancing.  It also and perhaps more importantly allows for consideration of the progress of coarser advance threads when successively distributing the workload of finer grids.  This ``dynamic load balancing'' allows adjustments to be made to finer level distributions to compensate for variations in progress made on ongoing coarser advances.

\begin{table}
\centering
\begin{tabular}{|c|l|}
\hline
\multicolumn{2}{|c|}{Variable List}  \\
\hline
$g_l^p$ & Current workload assigned to processor $p$ on level $l$ \\
$w_l^p$ & Amount of current workload $g_l^p$ completed \\
$c_l^p$ & Newly created child workload on level $l$ by parent grids on level $l-1$ on processor $p$ \\
$\overline{g_l}$ & Average of $g_l^p$ across processors \\
$\overline{w_l}$ & Average of $w_l^p$ across processors \\
$G_l$ & Sum of $g_l^p$ across processors.  Also equals sum of $c_l^p$ across processors. \\
$d_l^p$ & Desired workload assignments for processor $p$ and level $l$ \\
$\epsilon_l^p$ & Imbalance in $g_l^p$ ($g_l^p-\overline{g_l}$) \\
$\omega_l^p$ & Imbalance in $w_l^p$ ($w_l^p-\overline{w_l}$) \\
$\eta_l^p$ & Predicted remaining workload for entire AMR step for processor $p$ on level $l$ \\
$s_l$ & Current number of remaining steps on level $l$ within entire AMR step \\
$\delta_{p'}^p$ & Workload assigned to processor $p'$ of children created by grids on processor $p$ \\
\hline
\end{tabular}
\caption{List of variables and their definitions}
\label{variablelist}
\end{table}

  Here we give a brief example introducing terminology along the way.  Also see table \ref{variablelist} for a complete list of variables used below.  Consider a simulation run on three processors with two levels of refinement with a refinement ratio $R=2$ and let us assume (arbitrarily) that there are $G_0=72$ cells on level 0 distributed among the three processors as $g_0^p=[32,20,20]$.  Each processor may have more then one grid, but here we just count the total number of cells in all of the processor's grids.  After spawning the level 0 advance thread, each processor continues along the control thread until it eventually creates new child grids on level 1.  At this point these new child grids must be distributed.  Let's assume that at this point no work has yet been accomplished on the level 0 advance thread ($w_0^p=[0,0,0]$).  We also assume that the number of new child cells created by grids on each processor is $c_1^p=[32,8,8]$ for a total of ($G_1=48$) cells created on level 1 to be distributed across all three processors.  If we were independently balancing the workload on level 1 the desired distribution $d_1^p$ would be constant for each processor
\begin{equation}
d_1^p=\overline{g_1}
\end{equation}
where $\overline{g_1} = \overline{c_1} = \frac{G_1}{N}$ is the average new child workload.  For our example this would give a distribution of $d_1^p=[16,16,16]$.  There is however a workload imbalance on level 0 of $\epsilon_0^p=g_0^p-\overline{g_0}=[+8,-4,-4]$ that we would like to compensate for. In what follows we show how we can successively rewrite $d_1^p$ in new forms which allow us to better anticipate the best distribution to account for computation within and across levels.

If we completely compensate for the level 0 imbalance 
\begin{equation}
\label{verysimpledistributioneq}
d_1^p=\overline{g_1}-\epsilon_0^p
\end{equation}
 then the desired workload distribution for level 1 would be $[8,20,20]$.  If the level 1 workload is distributed that way, then while processor 1 waits for processors 2 \&  3 to complete their 20 level 1 updates, it should have time to complete 12 of its level 0 updates so $w_0^p=[12,0,0]$.  At this point the remaining workload on level 0 would be balanced at $g_0^p-w_0^p=[20,20,20]$.  However we need to take an additional step on level $1$.  Thus while the remaining level 0 workload is balanced, we see an opportunity to redefine $d_1^p$ in a way that accounts for this additional level 1 step.  We would like to redistribute the grids on level 1 such that they are balanced at $[16,16,16]$ before taking the second step to avoid processor idling.  This redistribution has an additional cost that can easily be avoided if the imbalance on level 0 is first weighted by the ratio of the number of level 0 steps $s_0$ to level 1 steps $s_1$ remaining.  During the first distribution of level 1 grids there are 2 remaining level 1 steps $s_1=2$ and one remaining level 0 step $s_0=1$.  Thus we redefine $d-1^p$ by modifying equation \ref{verysimpledistributioneq} to
\begin{equation}
\label{distributioneq2}
d_1^p=\overline{g_1}-\frac{s_0\epsilon_0^p}{s_1}
\end{equation}
This would give us $[16,16,16]-[+4,-2,-2]=[12,18,18]$. Now while processor 1 waits for processors 2 \& 3 to complete their 18 level 1 updates, it should have time to complete 6 of its level 0 updates effectively reducing the imbalance by a factor of 2.  After completing the first level 1 advance, the second distribution of level 1 grids must somehow take into account the coarser workload accomplished.  To do so we include a completed workload imbalance $\omega_l^p=w_l^p-\overline{w_l}=[4,-2,-2]$ into equation \ref{distributioneq2}.
\begin{equation}
\label{simpledistributioneq}
d_1^p=\overline{g_1}-\frac{s_0\epsilon_0^p-\omega_0^p}{s_1}
\end{equation}
Note that now with $s_1=1$ we get the same desired distribution of $[12,18,18]$ for the second level 1 step as for the first.  

We could have just used the remaining workload $h_0^p=g_0^p-w_0^p$ to calculate a remaining workload imbalance $\epsilon_0^p=h_0^p-\overline{h_0}$ and then used equation \ref{distributioneq2} but this would be the same as modifying equation \ref{simpledistributioneq} to be $d_1^p=\overline{g_1}-\frac{s_0\left(\epsilon_0^p-\omega_0^p\right)}{s_1}$.  Since in our example $\omega_0^p \neq 0$ only when $s_0 = 1$, it makes no difference, but when there are more then two levels of refinement this approach can lead to unnecessary shuffling of grids between processors even with static meshes.  For a more detailed analysis see \ref{staticdist}.

When there are multiple levels of refinement, the workloads on many coarser levels must be taken into account.  The generalization of equation \ref{simpledistributioneq} is given by
\begin{equation}
\label{distributioneq}
d_l^p=\overline{g_l}-\frac{\displaystyle \sum_{l'=0}^{l-1}{s_{l'}\epsilon_{l'}^p-\omega_{l'}^p}}{s_l}
\end{equation}

A more convenient form of equation \ref{distributioneq} can be obtained if we replace $\epsilon_{l}^p$ and $\omega_{l}^p$ with their expansions and then group terms by whether they involve averages across processors or local quantities on a particular processor.  We then arrive at the final general form of the distribution equation

\begin{equation}
\label{distributioneq3}
d_l^p=\overline{g_l}-\frac{\eta_{l-1}^p-\overline{\eta_{l-1}}}{s_l}
\end{equation}

where $\eta_{l-1}^p=\displaystyle \sum_{l'=0}^{l-1}{s_{l'}g_{l'}^p-w_{l'}^p}$.
Note that $\overline{\eta_{l-1}}$ is the predicted remaining work on all coarser levels and that if the actual distribution matches the desired distribution (ie $g_l^p=d_l^p$) then the total predicted remaining workload including the current level is

\begin{equation}
\label{longequality}
\eta_{l}^p=\eta_{l-1}^p+s_lg_l^p=\eta_{l-1}^p+s_ld_l^p=\eta_{l-1}^p+s_l\overline{g_l} - \eta_{l-1}^p+\overline{\eta_{l-1}} = s_l\overline{g_l}+\overline{\eta_{l-1}} = \overline{\eta_{l}}.  
\end{equation}

That is, each distribution attempts to equalize the predicted remaining work over the entire AMR advance.  It is not always possible to distribute a discrete set of grids perfectly so that $g_l^p=d_l^p$, however small differences on coarser levels can be corrected for on finer level distributions.  On the finest level the grids are artificially split to balance out all of the coarser level imbalances as described below.

\begin{figure}
 \caption{Example partitioning $\delta^{p'}_p$ of new level $l$ child workloads $c^{p'}_l$ onto desired workload shares $d^p_l$.  Shades of gray correspond to processor rank (0-3).}
 \centering
  \includegraphics[width=.8\textwidth]{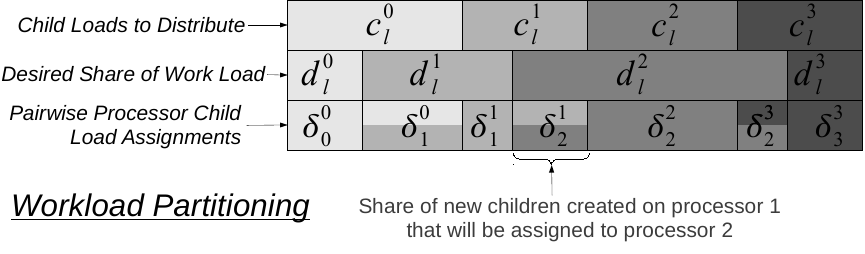} 
\label{partition}
\end{figure}

  Thus at each distribution of level $l$ grids, processors must collect the new child workloads $c_l^p$ to calculate $\overline{g_l}=\overline{c_l}$), and $\eta_l^p$ to calculate the desired workload distributions $d_l^p$.  Then the new child workloads $c_l^p$ are partitioned over $d_l^{p'}$ as shown in figure~\ref{partition} to give the desired workload $\delta_{p'}^p$ assigned to processor $p'$ of children created by grids on processor $p$.  Note that $\displaystyle\sum_{p'=1}^P\delta_{p'}^p=c_l^p$ and $\displaystyle\sum_{p=1}^P\delta_{p'}^p=d_l^{p'}$.  Each processor $p$ can then determine from which ''parent'' processors $p'$ it should expect to receive new child grids from ($\delta_{p}^{p'} > 0$) as well as which ``child'' processors $p''$ it should distribute new grids to ($\delta_{p''}^{p} > 0$)   Here the term child processor does not refer to the addition of new processors (unlike the addition/creation of new child grids) but only those existing processors which will receive child grids from other processors (referred to here as ``parents").

When distributing child grids among child processors, each processor $p$ will try to assign to each child processor $p'$ a collection of level $l$ grids with a combined workload $\delta_{p'}^p$.  However, arbitrarily partitioning a few discrete grids into exact sizes is often not possible, and the actual workload assigned $\delta_{p'}^p*$ to each child processor may be different from the desired workload $\delta_{p'}^p$.  As a result, the combined child workload from all of a processor's parent processors $g_{l}^p=\displaystyle\sum_{p'=1}^P\delta_{p'}^p*$ may be different from the desired workload $d_{l}^p$.  This results in some variation in $\eta_{l+1}^p$ between processors and a temporary predicted load imbalance.  However, if this variation is small, it can be corrected on finer level distributions or on the next round of level $l$ distributions without adversely effecting performance, provided that there remains enough computational work on coarser advance threads to buffer the imbalance.

\subsection{Grid Assignments}
\begin{figure}
 \caption{Plot showing Hilbert ordering of level 0 and level 1 cells.  Three colors correspond to processor containing the level 0 cells and the parent of the new level 1 cells.}
 \centering
  \includegraphics[width=1\textwidth]{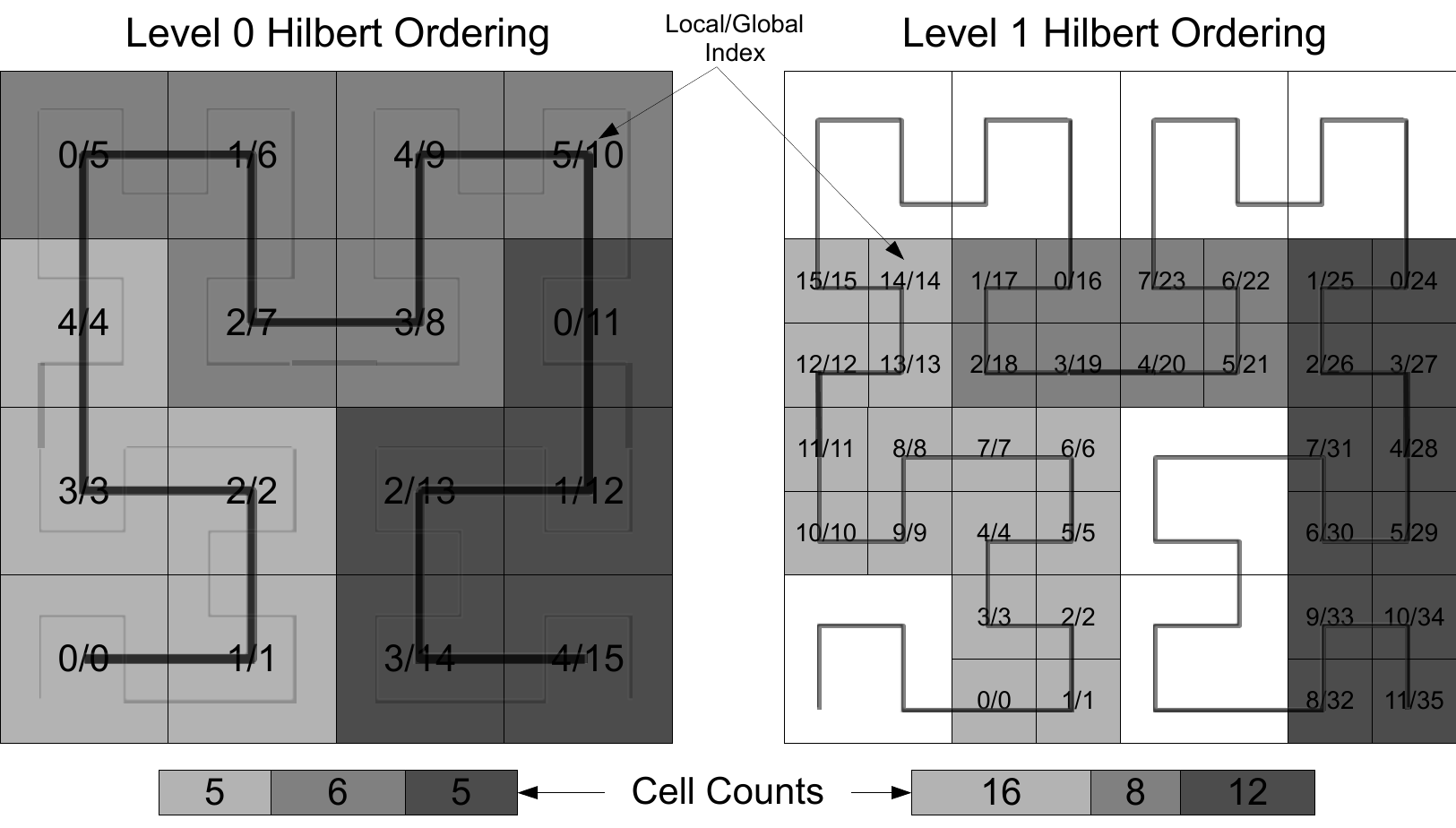} 
\label{hilbertfig}
\end{figure}

  Before discussing the details of the distribution of new refined grids, it is instructive to first consider the simpler problem of distributing newly refined cells as is the case for cell-based AMR.  After newly refined cells are created, a space filling curve is generally used to sort the new cells into a prescribed order based on their position along the curve \citep{MWK:IJPP01}. In addition the processors are ordered based on the topology of the network.  Then the ordered list of cells can be partitioned over the ordered list of processors in a manner similar to that used in constructing the $\delta_{p'}^p$ in figure~\ref{partition} except that now instead of globally distributing $c_l^p$ over $d_l^p$ to determine $\delta_{p'}^p$ we are locally distributing the individual workload of each child cell $\mathcal{C}^i_l$ over the child processor allocations $\delta_{p'}^p$ to determine the processor allocations for each individual child cell $\mathcal{D}_{p'}^i$.  Most cells will have only one nonzero allocation $\mathcal{D}_{p'}^i > 0$ and will be assigned to that processor $p'$ and if a cell has more than one nonzero allocation then the processor with the largest allocation could be used.  This ``rounding'' will only lead to load imbalances that are of order $\frac{N_{processors}}{N_{cells}}$.  This ordering attempts to keep communication between neighboring cells to a minimum.  For the partitioning to work, however, each processor needs to know the global sort index of each of its newly refined cells.  Sorting the new cells could be done using traditional sorting algorithms provided each processor was aware of all of the cells.  A better alternative, is to maintain a strict grouping of cells by sort order on each processor as shown in figure~\ref{hilbertfig}.  That is, a cell with a lower sort index will never be found on a higher numbered processor and vice versa.  Because of the fractal nature of many space filling curves like the Hilbert curves shown in figure~\ref{hilbertfig}, the ordering of any two child cells of different parent cells will be the same as that of its parents.  Therefore if the parent cells obey a strict grouping by sort order, the child cells will as well.  Processors therefore only need to determine the local sort order of their child cells.  They can then determine the global sort order if they know the new child cell counts of every other processor.  An example of these cell counts can be seen in figure~\ref{hilbertfig} as well as local and global index of each level 0 and level 1 cells.  The global index of each cell can be constructed by adding the cell counts of lower processors to the local index.  The fact that child cells will obey the same ordering of their parents can be used to locally sort child cells by first sorting the parent cells and then locally sorting the child cells among their siblings. 

  In cell based AMR each processor can usually partition its newly refined cells among ``child processors" with very little load imbalance since each processor typically has 1000's of cells to distribute over a few processors.  Not being able to split a cell might result in imbalances of few tenths of a percent.  This is not the case for patch based AMR.  Processors will still have 1000's of cells, but they will be grouped into a few grids of various sizes.  Distributing a few grids of various sizes among a few child processors with various workload assignments will in general lead to significant imbalances.  Often what is done in grid-based AMR is a more global distribution of all new child grids among all processors using a knapsack algorithm.  This type of distribution does not rely on a space-filling curves or any global ordering of new grids although it may take the communication costs into account.  This however, requires global knowledge of the AMR tree which can result in memory issues for simulations run on many processors.  In AstroBEAR, the AMR tree is distributed so a global knapsack type algorithm is not feasible.  Instead AstroBEAR uses an approach quite similar to that used for cell-based AMR.  AstroBEAR avoids the issues with load imbalances not by artificially fragmenting grids into small pieces, or by locally shuffling the grids in a local knapsack algorithm, but instead through the use of level threading described above in which load imbalances can be corrected for by subsequent distributions.  

This approach permits a strict grouping of grids on processors by their sort order.  This is true to at least to the extent that grids obey the same type of child order inheritance true for individual cells.  That is, while the order of two child cells will always be the same as that of their parent cells, it may not hold for grids.  Cells have a unique distance along the space filling curve, however grids contain many cells each with different distances.  A distance for the grid along the curve can be approximated by averaging that of all of its cells, or some subset of cells that appropriately samples the grid (ie the four corners or the center most subset).  However this fuzziness in grid distances can result in child grids occasionally being ordered differently then their respective parents.  However, this does not appear to be a major problem and would only result in slight increases in neighbor communication.  We note that instead of using a space-filling curve to order the grids, we could implement a local knapsack algorithm in which each processor distributes its own newly created children among its own child processors in an optimal fashion, however we have not investigated whether this improves performance compared to the ordering provided by a space-filling curve.

\subsection{Hilbert Splitting}
Occasionally splitting of a child grid into various pieces to accommodate its processor allocations $\mathcal{D}_{p'}^i$ is desired and there are two such instances in AstroBEAR.  First, since most of the workload resides on the finest level grids, imbalances can sometimes be too large to be compensated for by the coarser threads.  For this reason AstroBEAR always splits the finest level grids when necessary to achieve global load balancing.  In general the number of finest level grids split will be of the order of the number of processors.  Since the splitting occurs on the finest level, such fragmentation will not result in subsequent artificial fragmentation of higher level grids.  Second, since the base grid in AMR simulations can be quite large it can often contain more then a single processor's share of the entire workload.  For this reason the base grid is also split among the processors.  The algorithm for splitting begins by taking a single grid and its non-zero processor allocations $\mathcal{D}_{p'}^i > 0$ that were determined by locally partitioning the set of $\delta_{p'}^p$ over the individual child workloads $\mathcal{C}_l^i$.  This then gives the share of each individual child grid that should be assigned to each child processor.  It then attempts to split the grid into pieces each with a workload equal to the processor allocations $\mathcal{D}_{p'}^i$.  Additionally it attempts to construct the pieces so that they are properly ordered along the Hilbert curve.  This is done through a recursive bisection algorithm described below: 

\begin{enumerate}
\item  Divide the list of weights into two pieces as equal as possible.  For example if the list of weights were [.1 .2 .1 .3 .3] they would be split into [.1 .2 .1] and [.3 .3]
\item  Determine the two possible split points along each dimension that break the grid into two proportionate pieces (ie either [.4 .6] or [.6 .4]).
\item  For each of the possibilities evaluate the combined advance costs of the resulting grids.  This will in general favor splits along the longest grid dimension.
\item  If several possible splits have comparable combined advance costs then select the one that results in the largest difference in space-filling values between the two pieces with the correct sign.  
\item  If there were only two weights then we are done. Otherwise, continue to recursively bisect the left and right pieces.
\end{enumerate}

For a large fixed grid simulation, the above algorithm applied to a square base grid in 2D (or a cube in 3D) on $4^n$ processors ($8^n$ in 3D), will yield a distribution that traces out a level $n$ Hilbert curve.

 \section{Performance Results} \label{results}
 
  It is challenging to characterize the performance of an AMR code in general terms.  This is because the performance on any given problem will depend on many factors that are specific to that problem such as the number of levels of refinement, the number of cells on each level (filling fractions), their spatial distribution, and the number of processors used.  Here we give a few examples to demonstrate the complexity of the problem.  

  First consider a fixed grid 16x16 problem run on 5 processors.  Ideally the 16x16 grid could be split into 5 even rectangular pieces to distribute among 5 processors.  However, the best arrangement still leaves one processor with 55 cells which is 7\% more then the ideal average of 51.2 effectively limiting the performance to 93\%.  We could split the grid into more than 5 pieces so that the maximum number of cells on any processor is closer to the mean, but the additional overhead with having smaller grids would likely defeat the purpose.

The same problem however can be easily broken into 16 even pieces (each 4x4) to be perfectly distributed among 16 processors.  However, running the problem on 16 processors instead of 1 will not generally allow the problem to complete in 1/16 of the time for two primary reasons.  First, before each 4x4 grid can update, it most share ghost data with the surrounding grids and there is some network latency in transmitting ghost data between grids.  Additionally, while updating the 4x4 grid, calculations will be done using the ghost cells that will be repeated on the neighboring processors.  This redundant calculation is not required if there is only one 16x16 grid.  Thus updating 16 4x4 grids requires more calculations then updating 1 16x16 grid, so the CPU time will in general increase as more and more processors are used for the same problem even if the geometry of the problem allows for even distribution.

  Now consider the same 16x16 grid on 16 processors and lets assume that there are 16 refined regions (each 2x2) that just happen to lie in the center of each 4x4 grid.  These child grids will not be adjacent to other child grids and will only need to communicate with their parent grid.  If each child grid is distributed on the same processor containing the parent, then the communication pattern is the same as that of the fixed grid run and the performance should actually be better because each processor now has more work that can be done independently.  If however, instead of having 16 2x2 regions marked for refinement, the central 8x8 region was marked for refinement, then the communication pattern of the level 1 grids would be very similar to that of the level 0 although there would now be some additional communication between parent and child grids.  

  Thus the simplest and fairest test of an AMR code is to keep the number of cells per processor fixed as well as keeping the refined regions spatially connected.  Furthermore, if we keep the filling fractions at 1/4 for 2D (1/8 for 3D) then each level will have the same number of cells and the number of cells per level per processor will be a constant.  One still has the freedom in choosing the number of cells per leve per processor and this can have dramatic effects on the scaling.  Larger grids will in general allow for better scaling.  A fair test will choose a grid size that will allow for a simulation to complete in a reasonable wall time.  To that end we chose a resolution low enough so that a simulation on 10,000 cores would only take 8 hours per crossing time.  This works out to be $60^3$ cells per processor for the fixed grid run.  If we kept the base grid fixed and added 1 additional level of AMR, the simulation would take 3 times as long.  So we have to reduce the base grid size as we increase the maximum level of AMR to keep the overall walltime per crossing time fixed at 8 hours.  For 1 additional level of AMR this works out to be $46^3$ cells per processor per level, and $37^3$, $26^3$, \& $13^3$ for 2, 4, \& 8 levels of AMR respectively.  As the grids get smaller we expect the additional overhead related to ghost zones to reduce the overall speed - and the scaling will likely also suffer as well.  

\begin{figure}
 \caption{Results of weak scaling tests done on Kraken.  The number of cells per processor per level were set at $60^3,46^3,37^3,26^3,\&13^3$ for $0,1,2,4,\&8$ levels of AMR respectively.  Each plot shows the performance of the serial AMR approach, scheduled advances with level by level balancing as well as global balancing, and the threaded approach with level by level and global balancing.  The y-axis shows the number of cells updated per processor per second.}
 \centering
  \includegraphics[width=1\textwidth]{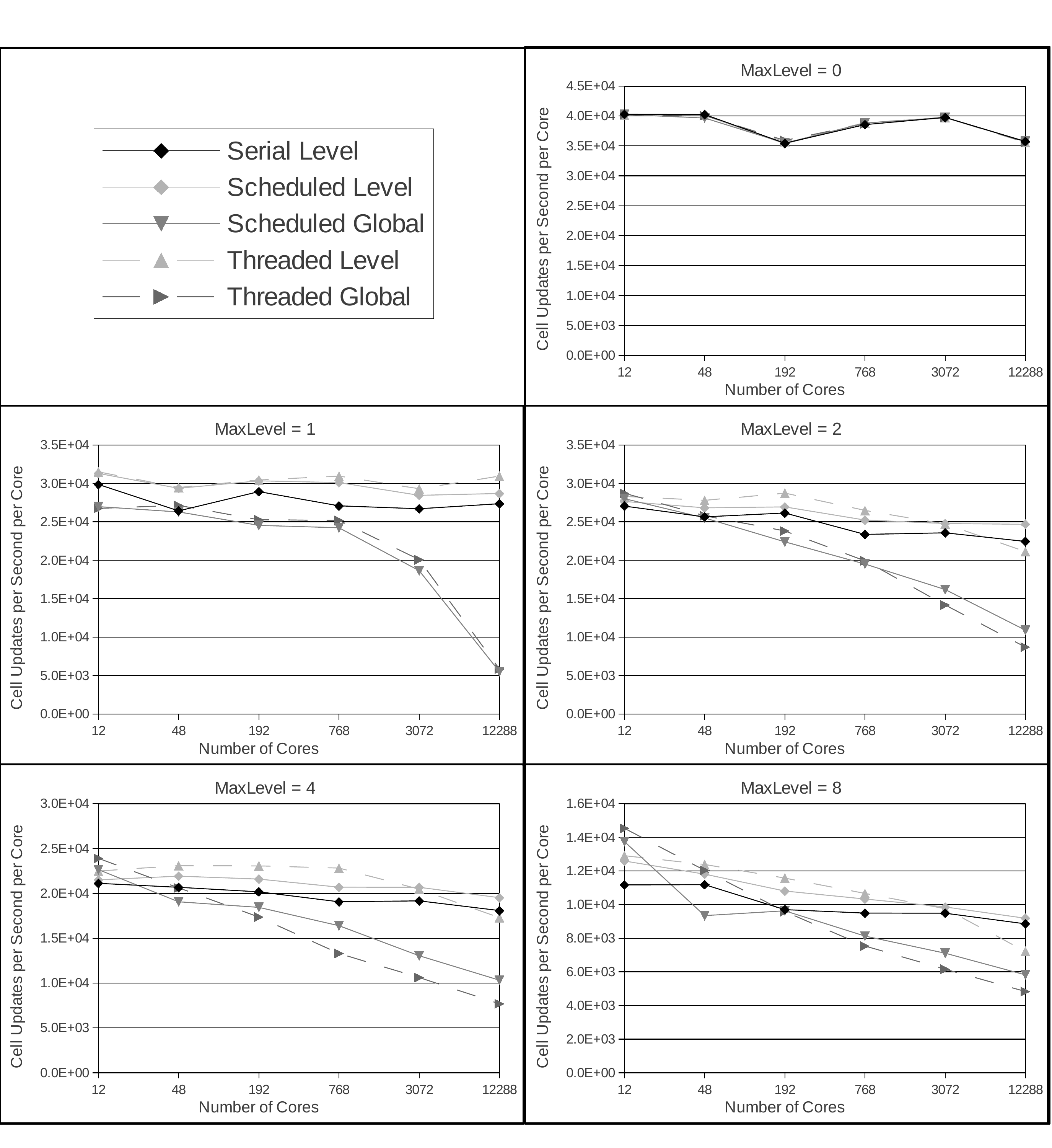} 
\label{WeakScaling}
\end{figure}

  Figure \ref{WeakScaling} shows the results of weak scaling done on Kraken.  For each level of AMR we ran simulations on 12, 48, 192, 768, 3072, \& 12288 cores using 5 different options for the load balancing/threading scheme.  For each simulation we calculated the total number of cell updates divided by the cpu time to get the efficiency of the code in actually solving the problem.  Other metrics such as the degree to which a processor remains idle, while interesting, are less relevant to the performance of actual simulations.  The solid black line shows the traditional serial AMR approach which in which each level is independently balanced out of necessity.  The solid light and dark grey line shows the results using a scheduling approach to threading in which processors schedule rendevouz times before switching to coarser advance threads.  The light line is for a simulation in which each level is balanced independently and the dark line is for a simulation that attempts to balance across levels or globally.  The dashed light and dark grey lines are for the simulations that actually use a threading library.  This allows for much finer grained switching between the control thread and the coarser advance threads.  Again the light line is for balancing within a level, and the dark line is for balancing across levels or globally.

  First there is a secular drop in performance for any number of cores in terms of cell updates per second per cpu as we increase the number of levels of AMR.  This is due to the decrease in the average grid size required as the number of AMR levels increases to keep the wall time constant.  The fixed grid run has typical grid sizes of $60^3$ while the run with 8 levels of AMR has typical grid sizes of only $13^3$.  Advancing cells within a $60^3$ grid can be done 4 times as efficiently as grids with only $13^3$ cells due to the overhead with regards to ghost zones etc.

 Next if we look at the serial AMR performance it turns out to be quite good with a worse case performance drop of only $\approx 20\%$ from 12 to 12288 cores.  We attribute this excellent performance to a fully parallelized tree and the presence of only a few scalar quantities that must be gathered across processors.  We also note in the upper right plot in figure \ref{WeakScaling} that for fixed grid there is no difference between the various approaches as expected since there is only 1 thread and no difference between local and global balancing. There is also a noticeable trend with regards to global vs. local balancing for both the scheduled and threaded approach.  The lower right plot of figure \ref{WeakScaling} shows most clearly this effect.  For 12 cores, which on Kraken are on a single node, the globally balanced approach out performs the locally balanced approach.  The globally balanced approach - while allowing for larger average grid sizes and therefore more efficient updating - reduces the degree to which various prolongation/restriction/synchonization operations are parallelized while also requiring more intercore communication.  When the cores reside on the same node, this communication can happen quickly and does not degrade the performance as much as the larger grid sizes improves the performance.  However by 48 cores the additional communication has negated any advantage of having larger grid sizes, and trying to keep grids as large as possible to avoid this sort of overhead does not seem to be effective on large numbers of cores.

  Finally if we compare the three level balanced approaches, we see that both the scheduled and the truly threaded approaches out perform the serial AMR approach by $5-20\%$.  There is a tendency for the fully threaded level balanced approach to drop in performance at 12288 cores, but this is likely due to the crude implementation of a non-blocking barrier.  The idea behind the threaded approach is that every communication is non-blocking and that while waiting for a message to complete coarser advance threads can be running while periodically checking to see if the communication had completed.  However global reductions cannot be done using non-blocking calls.  We dealt with this by writing a crude non-blocking barrier that could be executed before any global reductions and that would ensure every processor was ready to procede with the global mpi call.  The implementation involved every processor sending a ready signal to processor 0, followed by every processor waiting for a continue signal from processor 0.  This all-to-one followed by a one-to-all would likely cause significant performance degradation at 12288 cores.  There are more elegant solutions that could be implemented, such as scheduling these barriers or writing a more efficient branching communication pattern.

\section{Conclusion} \label{conclusion}
As discussed in the introduction Adaptive Mesh Refinement methods were designed to allow high resolution simulations to be carried out at low computational cost.  Highly parallel systems and their algorithms carried the same promise.  The parallelization of AMR algorithms is, however, not straight forward however and there have been a number of different approaches \citep{Paramesh2000, Nirvana2008, Enzo2004, Chombo, FTT}, to solve the problem.  While parallelization of a uniform mesh demands little communication between processors, AMR methods can demand considerable communication to maintain data consistency across the unstructured mesh as well as shuffling new grids from one processor to another to balance workload.  

In this paper we have described our attempt to design and implement a new strategy for AMR parallelization with an eye towards running codes on large machines with $>10^4$ processors. We have found that a threaded approach to the AMR algorithm significantly improves performance by allowing processors to remain busy while waiting for messages as well as to dynamically adjust distributions based on the progress of ongoing coarser grid updates.  We have also shown that a distributed tree algorithm significantly reduces the amount of memory required (and corresponding communication) for simulations run on many processors with modest grid sizes. 

We believe this threaded, distributed tree approach described in this paper holds considerable promise as a methodology for implementing AMR on ever-larger numbers of processors.  Future work will entail elaborating refinements to the methods as we experiment with its implementation on different classes of problems and machines. 
  
%\acknowledgments

\appendix
\section{Generalizing to arbitrary refinement ratios}
  In table \ref{geninheritance} we outline the procedure for inheriting neighbors, overlaps, from parent relationships for arbitrary number of refinement ratios.  And in table \ref{DistTreeAlg} we outline the communication steps for maintaining the distributed tree.
\begin{table}\sminy
\centering
\begin{tabular}{|>{\centering}p{.5in}|p{1.4in}|p{1.4in}|p{1.4in}|}
\hline
 Child iteration  & Node's child's neighbors & Node's child's preceding overlaps & Node's child's succeeding overlaps\\
\hline
 1 & Node's neighbors' 1$^{st}$ children & Node's preceding overlaps' R$^{th}$ children & Node's neighbor's 2$^{nd}$ children  \\
\hline
 2 & Node's neighbors' 2$^{nd}$ children & Node's neighbors' 1$^{st}$ children & Node's neighbors' 3$^{rd}$ children  \\
\hline
 3 & Node's neighbors' 3$^{rd}$ children & Node's neighbors' 2$^{nd}$ children & Node's neighbors' 4$^{th}$ children  \\ 
\hline
 ... & ... & ...  & ....  \\
\hline
 R-1 & Node's neighbors' R$^{th}$-1 children & Node's neighbors' R$^{th}$-2 children & Node's neighbors' R$^{th}$ children  \\
\hline
 R & Node's neighbors' R$^{th}$ children & Node's neighbors' R$^{th}$-1 children & Node's succeeding overlaps' 1$^{st}$ children \\
\hline
\end{tabular}
\caption{Generalized inheritance pattern for a refinement ratio of $R$}
\label{geninheritance}
\end{table}

\begin{table}
\begin{tabular}{| p{0.2in} | p{2.4in} | p{2.4in} |}
 
\hline
\multicolumn{3}{|c|}{First Step}  \\
\hline
 
  1 & \textit{Receive} new grids \& nodes along with their parents, neighbors, and \textbf{preceding} overlaps from parent processors & \textit{Receive} \textbf{succeeding} overlaps from parent processors \\
\hline
  2 & \multicolumn{2}{|p{4.8in}|}{Create new children and determine on which child processors they will go}\\
\hline
  3 & Determine which remote \textbf{preceding} nodes might have children that would overlap with its own children and \textit{send} the relevant children info & Determine which remote \textbf{succeeding} nodes might have created children that would overlap with its own children and \textit{send} the relevant children info \\
\hline
  4 & \multicolumn{2}{|p{4.8in}|}{Determine which remote \textbf{neighboring} nodes might have children that would neighbor its own children and \textit{send} the relevant children info} \\
\hline
  5 & \multicolumn{2}{|p{4.8in}|}{Determine which local \textbf{neighboring} nodes have neighboring children} \\
\hline
  6 & Determine which local \textbf{preceding} nodes have children that overlap with its own & Determine which local \textbf{succeeding} nodes have children that overlap with its own \\
\hline
  7 & \multicolumn{2}{|p{4.8in}|}{\textit{Receive} new children from remote \textbf{neighboring} nodes and determine which of the neighbors' children neighbors its own children} \\
\hline
  8 & \textit{Receive} children from remote \textbf{preceding} nodes and determine which of the nodes children overlaps with its own & \textit{Receive} children from remote \textbf{succeeding} nodes and determine which of the nodes children overlaps with its own. \\
\hline
  9 & For each remote child, \textit{send} the child grid's data as well as information about its parents, neighbors, \& \textbf{preceding} overlaps. & For each remote child, \textit{send} the child's \textbf{succeeding} overlaps. \\
\hline
\multicolumn{3}{|c|}{Successive Steps}  \\
\hline
  10 & \multicolumn{2}{|p{4.8in}|}{Create new children and determine on which child processor they will go} \\
\hline
  11 & \multicolumn{2}{|p{4.8in}|}{Determine which remote \textbf{neighboring} nodes might have old/new children that would overlap/neighbor its own new children and \textit{send} the relevant children info}  \\
\hline
  12 & \multicolumn{2}{|p{4.8in}|}{Determine which local \textbf{neighboring} nodes might have old/new children that would overlap/neighbor its own new children}  \\
\hline
  13 & \multicolumn{2}{|p{4.8in}|}{\textit{Receive} new children from remote \textbf{neighboring} nodes and determine which of the neighbors' children neighbors/overlaps its new/old children} \\
\hline
  14 & \multicolumn{2}{|p{4.8in}|}{For each \textbf{new} remote child, \textit{send} the child's information, and the information about its parent, neighbors, \& \textbf{preceding} overlaps.} \\
\hline
  15 & \multicolumn{2}{|p{4.8in}|}{For each \textbf{old} remote child, \textit{send} the child's \textbf{succeeding} overlaps.} \\
\hline
\hline
\end{tabular}
 \caption{The split rows denote actions taken by the current iteration of nodes (left) and the old iteration of nodes (right).  Otherwise the actions are taken only by the current iteration of nodes.}
\label{DistTreeAlg}
\end{table}

\section{Static Distributions for a Static Mesh}\label{staticdist}
  Consider what happens if the mesh is static so that the average workload per level $\overline{g_l}$ remains constant.  Initially there will be a desired set of distributions $d_l^p$ that will result in an actual set of distributions $g_l^p$.  On the finest level they will agree, but on coarser levels there will be some difference between $d_l^p$ and $g_l^p$.  It is desirable that each successive desired distribution should match the original so that each successive actual distributions will also remain constant and grids will not be unnecessarily shuffled around.  For example, consider the desired distribution of the highest level $L$ after the first level $L$ advance.  Assuming the highest level workload follows the desired workload, $g_L^p=d_L^p=\overline{g_L}-\frac{\eta_L^p-\overline{\eta_L}}{s_L}$.  The amount of time each processor will have waiting for every other processor to complete the level $L$ advance is just $\displaystyle \max_p(g_L^p)-g_L^p$.  If this time is spent advancing coarser grids, then the new sum of work done on coarser grids $\displaystyle \sum_{l'=0}^{L-1}{w_l^{p*}} = \displaystyle \sum_{l'=0}^{L-1}{w_l^p} + \left [\max_p(g_L^p)-g_L^p \right ]$.  This gives a new predicted remaining work $\eta_L^{p*} = \eta_L^p-\left [\displaystyle \max_p(g_L^p)-g_L^p \right]$ and a new desired distribution 

\begin{eqnarray}
d_L^{p*} &=& \overline{g_L}-\frac{\eta_L^{p*}-\overline{\eta_L^{p*}}}{s_L*} = \overline{g_L}-\frac{\eta_L^p-\left ( \displaystyle \max_p(g_L^p)-g_L^p \right)-\left [\overline{\eta_L}-\left( \displaystyle \max_p(g_L^p)-\overline{g_L}\right )\right]}{s_L-1} \nonumber \\
&=& \overline{g_L}-\frac{\eta_L^p-\overline{\eta_L} + g_L^p-\overline{g_L}}{s_L-1}=\overline{g_L}-\frac{\eta_L^{p}-\overline{\eta_L^{p}}}{s_L} = d_L^p \nonumber \\
\end{eqnarray}

More generally we can consider the distribution of any given level $l$ after completing $c_{l}$ steps leaving $s_{l}=2^l-c_{l}$ remaining steps.  
\begin{eqnarray}
d_l^p=\overline{g_l}-\frac{\eta_l^p-\overline{\eta_l}}{s_{l}} \nonumber
\end{eqnarray}
If $\eta_l^p-\overline{\eta_l} \propto s_{l}$, then the desired distribution will remain fixed.
\begin{eqnarray}
\eta_l^p-\overline{\eta_l}&=&\left (\displaystyle \sum_{l'=0}^{l-1}{s_{l'}g_{l'}^p-w_{l'}^p} \right ) - \left( \displaystyle \sum_{l'=0}^{l-1}{s_{l'}\overline{g_{l'}}-\overline{w_{l'}}} \right ) \nonumber \\
\end{eqnarray}
The workload completed on the coarser grids will be the difference between the maximum work completed on all of the previous grids by any processor and the local work completed on all of the previous grids.  This then gives
\begin{eqnarray}
\displaystyle\sum_{l'=0}^{l-1}w_{l'}^p=\displaystyle \max_p{\left [\sum_{l'=0}^L(2^{l'}-s_{l'})g_{l'} \right ]} - \sum_{l'=0}^L(2^{l'}-s_{l'})g_{l'}^p \nonumber \\
\end{eqnarray}
and we have
\begin{eqnarray}
\eta_l^p-\overline{\eta_l}&=&\left (\displaystyle \sum_{l'=0}^{l-1}{s_{l'}g_{l'}^p} + \sum_{l'=0}^L{(2^{l'}-s_{l'})g_{l'}^p} \right ) - \left( \displaystyle \sum_{l'=0}^{l-1}{s_{l'}\overline{g_{l'}}+   \sum_{l'=0}^L{(2^{l'}-s_{l'})\overline{g_{l'}}}} \right )\nonumber \\
&=&\displaystyle \sum_{l'=0}^{L}{2^{l'}\left(g_{l'}^p-\overline{g_{l'}}\right)}+\sum_{l'=l}^L{s_{l'}\left(\overline{g_{l'}}-g_{l'}^p\right)} \nonumber \\
\end{eqnarray}

Now $\displaystyle \sum_{l'=0}^L 2^{l'}g_{l'}^p = \eta_{L+1}^p$ it the initial entire predicted workload and is balanced by the highest level distribution so the value on any processor equals the average and the first term cancels.  The second term can be re-written by recognizing that at the moment we are distributing level $l$, for $l'>=l$, $s_{l'}=2^{l'-l}s_{l}$.  This gives
\begin{eqnarray}
\eta_l^p-\overline{\eta_l}&=&s_{l}\sum_{l'=l}^L{2^{l'-l}\left(\overline{g_{l'}}-g_{l'}^p\right)} \propto s_l\nonumber \\
\end{eqnarray}
Therefore if the mesh is static, the distribution will remain static as well.

%\end{appendix}

\end{document}